\documentclass[epj]{svjour}
\usepackage{graphics}
\usepackage{color}

\usepackage{cancel}
\begin{document}

\title{
Numerical analysis of electronic conductivity in graphene with resonant adsorbates: 
comparison of monolayer and Bernal bilayer 
}

\titlerunning{Numerical analysis of conductivity in bilayer graphene}
\authorrunning{A. Missaoui {\it et al.}}

\author{Ahmed Missaoui\inst{1},
 Jouda Jemaa {Khabthani}\inst{2},
  Nejm-Eddine {Jaidane}\inst{1},
  Didier {Mayou}\inst{3,4} \and
  Guy {Trambly de Laissardi\`ere}\inst{5} 
}                     
\institute{Laboratoire de Spectroscopie Atomique Mol\'eculaire et Applications, 
D\'epartement de Physique, Facult\'e des Sciences de Tunis, Universit\'e Tunis El Manar,  Campus Universitaire 1060 Tunis, Tunisia
\and Laboratoire de Physique de la mati\`ere condens\'ee,
D\'epartement de Physique, Facult\'e des Sciences de Tunis, Universit\'e Tunis El Manar,  Campus Universitaire 1060 Tunis, Tunisia
\and Universit\'e Grenoble Alpes, Inst NEEL, 38042 Grenoble, France
\and CNRS, Inst NEEL, 38042 Grenoble, France
\and Laboratoire de Physique th\'eorique et Mod\'elisation, CNRS and 
Universit\'e de Cergy-Pontoise, 
95302 Cergy-Pontoise, France}
%
\date{January 27, 2017}
%
\abstract{ 
We describe the electronic conductivity, as a function of the Fermi energy, in the Bernal bilayer graphene (BLG) in presence of a random distribution of vacancies that simulate resonant adsorbates. We compare it to monolayer (MLG) with the same defect concentrations. These transport properties are related  to the values of  fundamental length scales such as the elastic mean free path $L_{e}$,  the localization length $\xi$  and the inelastic mean free path $L_{i}$. 
Usually the later, which reflect the effect of inelastic scattering by phonons, strongly depends on temperature $T$. In  BLG an additional characteristic distance $l_1$ exists which is the typical traveling distance  between two interlayer hopping events. We find that when the concentration of defects is smaller than 1\%--2\%, one has $l_1 \le L_e \ll \xi$ and the BLG has transport properties that differ from those of the  MLG  independently of  $L_{i}(T)$. Whereas for larger concentration of defects $L_{e} < l_1 \ll \xi $, and depending on $L_{i}(T)$, the transport in the BLG can be equivalent (or not)  to that of two decoupled MLG. 
{
We compare two tight-binding model Hamiltonians with and without hopping beyond the nearest neighbors.
}
} 
\maketitle
\section{Introduction}
\label{intro}
Graphene consists of a monolayer (MLG) carbon atoms, with $sp^2$ hybridization, forming a 2D honeycomb lattice, with two equivalent atoms --atom A and atom B-- in a unit cell. 
Linear dispersion relation of the $p_z$ electron states close to the Fermi energy induces many fascinating transport properties which give rise to potential device applications \cite{Wallace47,Berger06,Katsnelson06,Castro09_RevModPhys}. 
Few-layer graphene also present unusual properties. In particular the Bernal bilayer graphene (BLG) with $AB$ stacking, as in graphite, 
breaks the atom A / atom B symmetry and leads to quadratic dispersion relation \cite{Latil06,Castro07,MacdoABC,Varchon_prb08,Ohta06,Brihuega08,DasSarma11,McCann13,Ulstrup14}.

Electronic transport is sensitive to static defects which are for example screened charged impurities, or local defects like vacancies or adsorbates, 
{
(hydrogen, adatoms or admolecules,
chemically bound to one carbon atom of the surface of graphene layer).
} 
Theoretical studies of the effects introduced by the adsorbates on the conductivity  has been done for MLG 
(Refs. \cite{V.M.Pereira,Robinson08,Yuan10b,Yuan12,Lherbier12,Roche12,Roche13,Kretinin13,Trambly11,Trambly13,Trambly14,Zhao15} and Refs. therein),
and  for BLG \cite{Koshino06,Adam08,Koshino09,Yuan10,Gonzalez10,Ferreira11,VanTuan16,McCann13}.
Most of them consider a standard Hamiltonian that takes only into account the  hopping between the nearest neighbors orbitals.
{
Yet some studies show the importance of hopping beyond nearest neighbors on electronic structure and transport properties \cite{Castro09_RevModPhys,V.M.Pereira,Trambly14,Zhao15,VanTuan16}.
}
In this paper, we present electronic properties of MLG and BLG obtained by two tight-binding (TB) models: 
the standard model with nearest neighbor only (TB1) and a TB model including the effect of the hopping beyond nearest neighbors (TB2).
This second model, which is more realistic, predicts some differences in the transport at energies close to the resonant energy of scatters. 

{
We consider local defects, such as adsorbates or vacancies, that are resonant scatters.  
Local defects tend to scatter electrons in an isotropic way for each valley and lead also to strong intervalley scattering. The $T$ matrix of a local defect usually depends strongly on the energy. In the case of simple vacancies or adsorbates (atoms or molecules) that create one covalent bound with a carbon atoms of MLG (BLG), the $T$ matrix diverges at the energy $E_{MG}$ 
(with TB1 model $E_{MG}=E_D=0$).
For this reason, theses scatters are called resonant scatters.
}
The adsorbate is simulated by a simple vacancy in the plane of $p_z$ orbital 
as usually done \cite{Castro09_RevModPhys,McCann13,Roche13,Trambly13,VanTuan16}. 
Indeed the covalent bonding between the adsorbate and the carbon atom of graphene to which it is linked, 
eliminates the $p_z$ orbital from the relevant energy window. The scatterers are distributed randomly in both planes and with the same concentration in both planes. 
{
We consider  here that the up and down spins are degenerate i.e. we deal with  a paramagnetic state. Indeed the existence and the effect of a magnetic state for various adsorbates or vacancies is still debated \cite{Nair12,Scopel16}.  Let us emphasize that  in the case of a magnetic state  the up and down spins  give two different contributions to the conductivity but the individual contribution of each spin can be analyzed from the results discussed here. 
}

We first determined the density of states (DOS) in disordered MLG and BLG in presence of static scatterers (vacancies) with various concentrations $c$ from 0.5\% to 10\%. 
{
Elastic mean free path $L_e$, which depends on the distribution of scatters and on energy $E$, is also computed.
From diffusive properties of wave packet in the structure, the electrical conductivity $\sigma$ 
is computed versus $E$ and the inelastic mean free path $L_i$. $L_i$ due to electron-phonons interaction or magnetic fied.
Roughly speaking, large $L_i$ correspond to low temperature limit and small $L_i$ to room temperature.
The numerical method used takes into account all quantum effects.
We show that difference between MLG and BLG is explained by considering the average distance $l_1$ over which a charge carrier travels in  a layer between two interlayer hoppings \cite{McCann13,Snyman07}. 
As explained in this paper, $l_1  \simeq$\,2nm and it is almost independent on the concentration of defects. 
Indeed BLG has some similar properties to MLG for small $L_i$ and large $c$, $i.e.$ when $L_e \simeq L_i < l_1$. 
But for small $c$ values, $L_e \simeq L_i \ge l_1$,  the effects of interlayer hopping affect the electronic properties of BLG with respect to MLG case. 
In this case, the conductivity of the BLG varies with $c$ like in usual metals where DOS is finite whereas MLG behaves like a semi-metals with Dirac electrons.
Moreover, different regimes of transport in BLG are found depending on the values of the energy, 
like in MLG \cite{Roche13,Trambly13,Zhao15}.
}

{
Finally the localization length $\xi$ is computed and we study the localization regime (which can be observed experimentally in a very low temperature regime, i.e. large $L_i$). 
For the concentrations studied here, $l_1 < \xi$, which means that in this localization regime the coupling between the two layers plays always a significant role. Thus behavior of the BLG and MLG are different and  localization length is larger in the BLG.
}

In Sect. \ref{secElecStruc}, the two tight binding (TB) models used are described and the corresponding density of states are discussed. 
Transport properties are presented in Sect. \ref{secTransp}. We first describe rapidly the computational method and the relevant lengths to analyze conductivity in BLG (Sect. \ref{secCompt}). Then elastic scattering length $L_e$ (Sect. \ref{secLe}) and microscopic conductivity $\sigma_M$ (Sect. \ref{secSigmaM}) are presented.
Finally quantum interference corrections to the conductivity and localization length $\xi$  are analyzed (Sect. \ref{secLocalization}).  
Sect. 4 is devoted to concluding remarks. 

\section{Electronic structure}
\label{secElecStruc}
\subsection{Tight binding Hamiltonian models}
\begin{figure}
\begin{center}
\rotatebox{-90}{\resizebox{0.22\textwidth}{!}{%
  \includegraphics{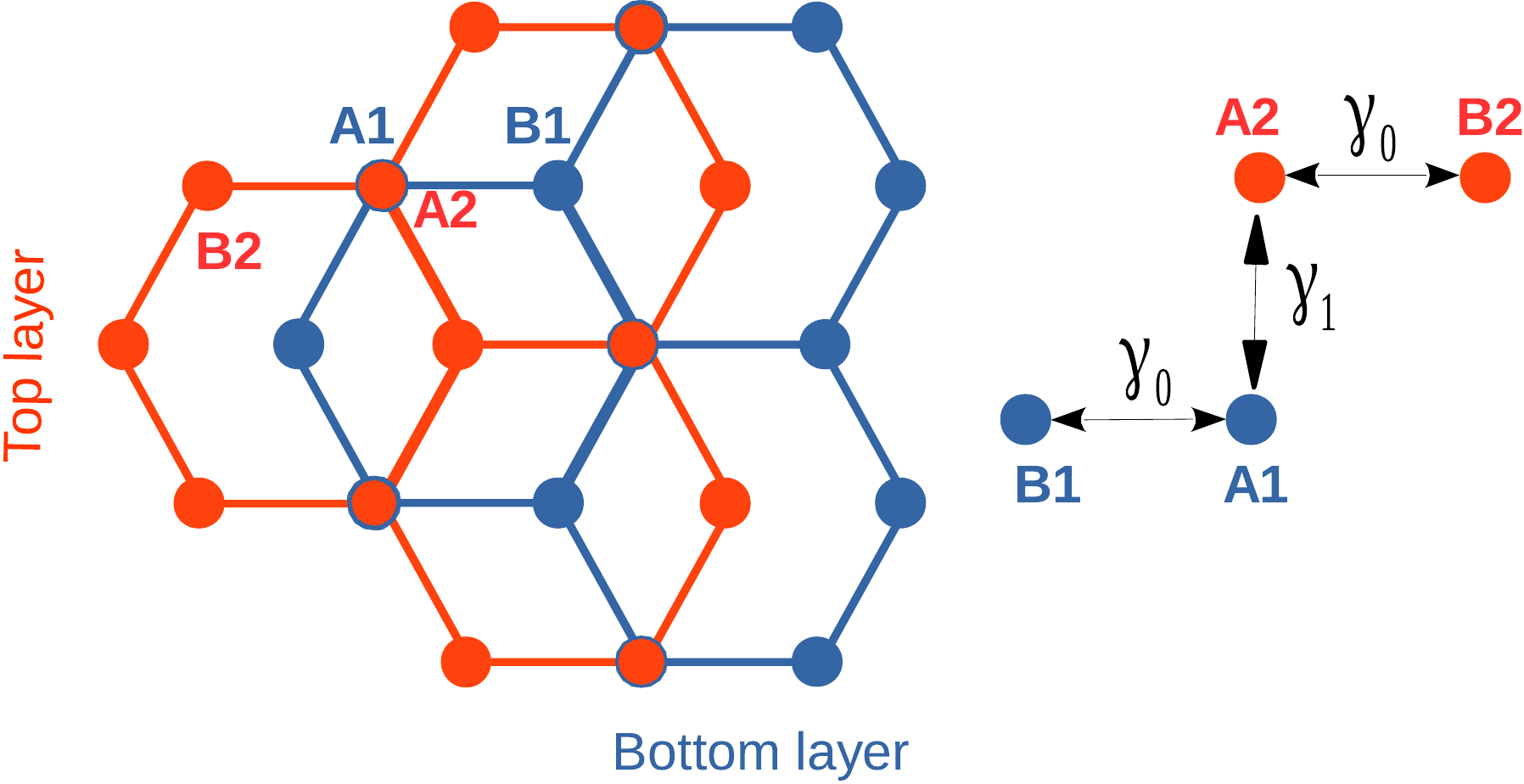}
}}
\caption{\label{struc} Planar view of the crystal structure of AB stacked bilayer graphene. Atoms $A_{1}$ and $B_{1}$ on the lower layer are shown as blue circles. $A_{2}$, $B_{2}$ on the upper layer are orange circles.}
\end{center}
\end{figure}

BLG can be considered as two coupled MLGs with the top layer shifted by a carbon bond from the bottom layer. Consequently, BLG consists of four atoms in its unit cell, two carbons A$_1$, B$_1$ from the unit cell in the bottom layer and A$_2$, B$_2$ in the the top layer where A$_2$ sits at the top of A$_1$ (Fig. \ref{struc}). 
We used a tight binding scheme (TB) \cite{Trambly12}.
Only p$_z$ orbitals are taken into account since we studied the electronic properties around the Fermi energy  level. Interlayer interactions are not restricted to the $pp\sigma$ terms but $pp\pi$ terms have also to be introduced. The Hamiltonian has the form:
\begin{equation}
\hat{H} =\sum_{i} \epsilon_{i}|i\rangle\langle i| + \sum_{(i,j)}t_{ij}|i\rangle\langle j| ,
\end{equation}
where $i$ is the orbital located at $\bf{r}_{i}$ and 
the sum runs over all neighboring $i$, $j$ sites. 
The energy on the site is  taken equal to zero for first nearest neighbor model. $t_{ij}$ is the hopping element matrix between site $i$ and site $j$, computed from the Slater-Koster parameters,
\begin{eqnarray} 
t_{ij} 
~=~   n_c^2 V_{pp\sigma}(r_{ij}) ~+~ (1 - n_c^2) V_{pp\pi}(r_{ij}),
\end{eqnarray}
where $V_{pp\sigma}$ and $V_{pp\pi}$ depend from the distance $r_{ij}$ and $n_{c}$ is the cosines direction along the $Oz$ axis. It is either equal to zero or to a constant because the two graphene layers have been kept flat in our model. 
The Slater Koster parameters are exponentially decaying function of the distance: 
\begin{eqnarray}
\label{slater}
V_{pp\pi}(r_{ij})    &=&-\gamma_0 \exp { \left( q_{\pi}    \left(1-\frac{r_{ij}}{a} \right) \right)} , \label{eq:tb0}\\
V_{pp\sigma}(r_{ij}) &=& \gamma_1 \exp { \left( q_{\sigma} \left(1-\frac{r_{ij}}{a_1} \right) \right)} .
\label{eq:tb1}
\end{eqnarray}
It allows, according to the value of $q_{\pi}$, to take into account both first neighbors or second neighbors. 
$a$ is the nearest neighbor distance within a layer, $a=1.418$\,{\rm \AA}, and  $a_{1}$ is the interlayer distance, $a_1=3.349$\,{\rm \AA}. First neighbors interaction in a plane is characterized by the commonly used value $\gamma_{0}=2.7$\,eV and the second neighbors interaction $\gamma_{0}^{'}$ is set to $0.1\gamma_{0}$. 
The ratio ${q_{\pi}}/{a}$ in Eq. \ref{slater} is fixed by the value of the $\gamma_{0}^{'}$. The interlayer coupling between two $p_{z}$ orbitals in 
$\pi$ configuration is $\gamma_{1}$,  $\gamma_{1}=0.48$\,eV, and it is fixed to obtain a good fit with  {\it ab initio} calculation around Dirac energy in AA stacking and AB bernal stacking \cite{Trambly12}. 
It is worth to note that we choose the same decay coefficient  for $V_{pp\pi}$ and $V_{pp\sigma}$:
\begin{eqnarray}
\frac{q_{\sigma}}{a_1} ~=~ \frac{q_\pi}{a} 
~=~ \frac{{\log} \left(\gamma_0/\gamma_0' \right)}{a' - a}
~=~ 2.218{\rm \AA}^{-1},
\end{eqnarray}
with $a'=2.456$\,{\rm \AA} the distance between second neighbors in a plane. All $p_{z}$ orbitals have the same on-site energy $\epsilon_{i}$ in two planes. 
In order to obtain a Dirac energy $E_D$ equal to zero in MLG, one fixes $\epsilon_{i}$ equal to $-0.78$\,eV for TB model with hopping beyond first neighbors. 
This is necessary because hopping beyond first neighbors breaks the electron/hole symmetry and then shifts $E_{\rm D}$ value 
{\cite{Castro09_RevModPhys,Kretinin13}}.
We have used this model Hamiltonian in our previous work \cite{Trambly12,Trambly10,Trambly16} to study the electronic structure rotated graphene bilayer. 
In order to analyze the effect of hopping beyond the first-neighbor distances, we consider also the simplest TB model (TB1) with first-neighbor hopping only and the same parameters than the complete TB2 model described above.

As explained in the introduction we consider that resonant adsorbates --simple atoms or molecules such as H, OH, CH$_3$--  create a covalent bond with some carbon atoms of the BLG. To simulate this covalent bond, we assume that the $p_{z}$ orbital of the carbon, that is just below the adsorbate, is removed. In our calculations the mono-vacancies are distributed at random between the both planes with a finite concentration $c$.

\subsection{Density of states}
\label{secDOS}

\begin{figure}
\begin{center}
\resizebox{0.45\textwidth}{!}{%
  \includegraphics{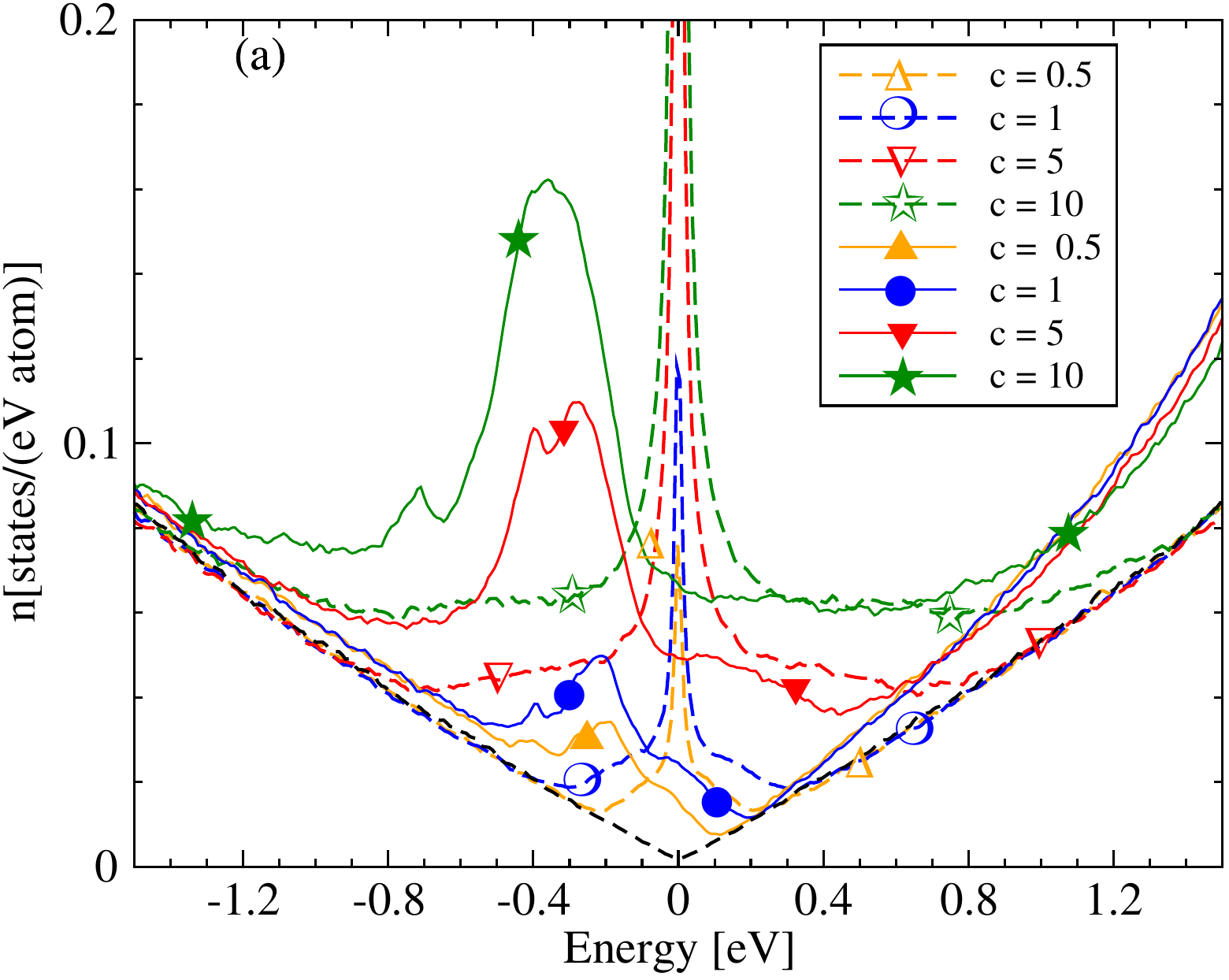}
}
\resizebox{0.45\textwidth}{!}{%
  \includegraphics{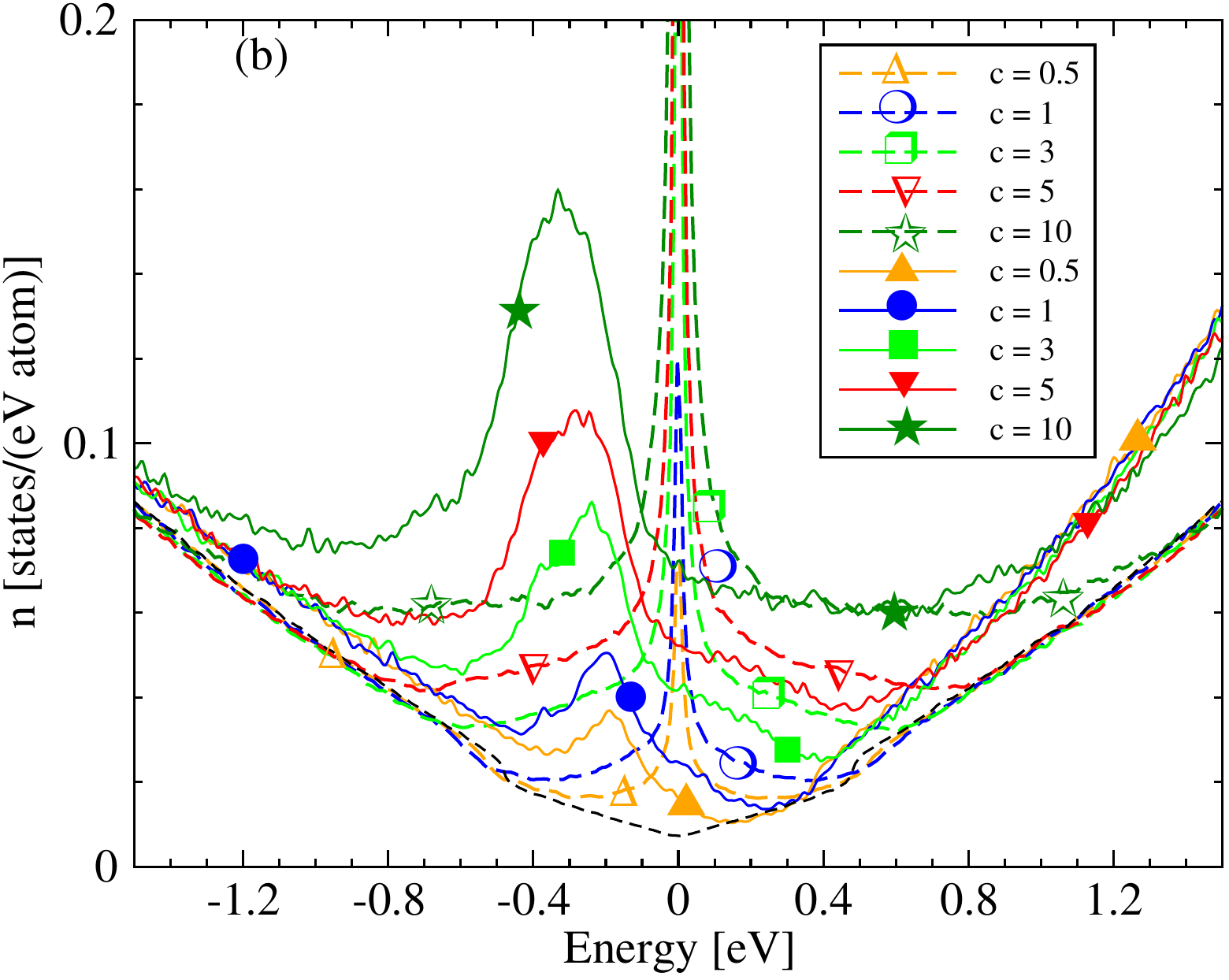}
}
\caption{\label{Fig2}
(color online)
Total density of states (DOS) $n$  versus energy $E$, for concentrations $c$ [\%] of resonant adsorbates (vacancies):
(dashed line) TB1 (first neighbor hopping only),
(solid line) TB2 (with hopping beyond first neighbor); (a) for MLG, (b) for BLG. 
The black dashed curve is pristine MLG (BLG), $i.e.$ $c=0$, calculated by TB1. 
}
\end{center}
\end{figure}

Figure $\ref{Fig2}$(b) shows the total density of states (total DOS) $n(E)$ in BLG for different concentrations $c = 0.5\%$ to $10\%$ of defects randomly distributed.
For comparison, the DOSs of MLG with same TB Hamiltonian 
{
\cite{Trambly13,Trambly14} 
}
are  also shown (Fig. \ref{Fig2}(a)).
Without defects, $c=0$, BLG DOS and MLG DOS differ for energies $E$ such as 
$-\gamma_1 < E < \gamma_1$ \cite{McCann13}. For small $c$ concentrations, $c< \sim 1\%$, this distinction is still observed. 
But remarkably, MLG and BLG have very similar DOS, for concentration of defects $c$ larger than $\sim 1\%$. 

With TB1 model (first-neighbor hopping only), states occur at energy $E_{\rm MG}=0$. This is reminiscent of the midgap state produced by asymmetry between the number of atoms A and  B in monolayer graphene \cite{P.W.Brouwer,V.M.Pereira}. 
{
In agreement with previous findings for MLG \cite{Castro09_RevModPhys,V.M.Pereira,Yuan10b,Roche13} and BLG \cite{McCann13,Yuan10,VanTuan16},
}
for large values of energies  $E$ the DOS is weakly affected by the presence of disorder. Finally near the Dirac energy there is an intermediate regime where the pseudo-gap is filled (``plateau''). 

With the TB2 model including all neighbors, the midgap state is no longer at $E=0$, but it becomes a broad peak at negative energy $E_{\rm MG}$. 
$E_{\rm MG}$ value varies from $E_{\rm  MG}\simeq-0.2\,eV$ to $E_{\rm MG} \simeq-0.3\,eV$ when defect concentration increases from $c=0.5\%$ to $10\%$, like in monolayer graphene (Fig. \ref{Fig2}(a)) 
{
\cite{Castro09_RevModPhys,V.M.Pereira,Trambly14,Zhao15}. 
}
In the following we distinguish  three cases according to energy values:
\begin{itemize}
\item[$(i)$] Sufficiently large energies, BLG DOS and MLG DOS are similar and they are not strongly modified by the presence of resonant defects. 
\item[$(ii)$] Energies in the ``plateau'' due to vacancies but not in the midgap states $E \neq E_{\rm MG}$.
\item[($iii$)] Energies in the midgap states, $i.e.$ $E=E_{\rm MG}=0$ for TB1 (with first-neighbor hoppings only) and $E\simeq E_{\rm MG}$ for TB2 (with hopping beyond first neighbors). 
\end{itemize}
These three cases correspond to different transport 
regimes.

\section{Transport properties}
\label{secTransp}

\subsection{Computational method and relevant lengths}
\label{secCompt}

In the framework of the Kubo-Greenwood formula for electronic transport properties, 
the quantum diffusion coefficient $D$ (diffusivity) and conductivity $\sigma$ are computed by using the polynomial expansion method, developed by Mayou, Khanna, Roche and Triozon \cite{Mayou88,Mayou95,Roche97,Roche99,Triozon02}.  
This numerical approach
allows very efficient calculations by recursion (Lanczos algorithm) in real-space which take into account all quantum effects.
It has been used to study quantum transport in disordered graphene, chemically doped graphene and bilayer (see Refs. \cite{Lherbier12,Roche12,Roche13,Trambly11,Trambly13,Trambly14,VanTuan16} and Refs. therein). 
Our calculations are performed on sample containing up to a few $10^{7}$ carbon atoms, which corresponds to typical sizes of about one micrometer square and allows to study systems with elastic mean-free length of the order of few hundred nanometers.

Elastic scattering events are taken into account in the Hamiltonian, but effects of inelastic scattering by phonons at temperature $T$ are not included in the Hamiltonian. To consider the inelastic processes, we introduce an inelastic scattering time $\tau_{i}(T)$ beyond which the propagation becomes diffusive due to the destruction of coherence by these inelastic processes. The effect of a magnetic field on the electron propagation is not included  directly in the TB model, but a magnetic field $B$ can have also a similar 
{
incoherent
}
dephasing effect.
This dephasing effect  occurs on a length $L_{i}(B)$ such that the flux of the magnetic field enclosed in the disk of radius $L_{i}(B)$ is equal to the flux quantum $h/e$, i.e. $L_{i}(B)\simeq \sqrt{h/eB}$. 
We treat these two dephasing effects in a phenomenological way through a Relaxation Time Approximation (RTA) as described here after.
In the RTA, the conductivity  along the $x$-axis is given by, \cite{Trambly13}
\begin{eqnarray}
\sigma(E_{F},\tau_{i}) &=& e^{2}n(E_{F})D(E_{F},\tau_{i}) , \\
D(E_{F},\tau_{i}) &=& \frac{L_{i}^{2}(E_{F},\tau_{i})}{2\tau_{i}} , 
\end{eqnarray}
where $E_F$ is the Fermi energy, $n(E_{F})$ is the density of states (DOS)  and $L_{i}$ is the inelastic mean-free path conductivity  along the $x$-axis. $L_{i}(E_F,\tau_{i})$ is the typical distance of propagation during the time interval $\tau_{i}$ for electrons at energy $E$.

\begin{figure}
\begin{center}
\resizebox{0.43\textwidth}{!}{%
  \includegraphics{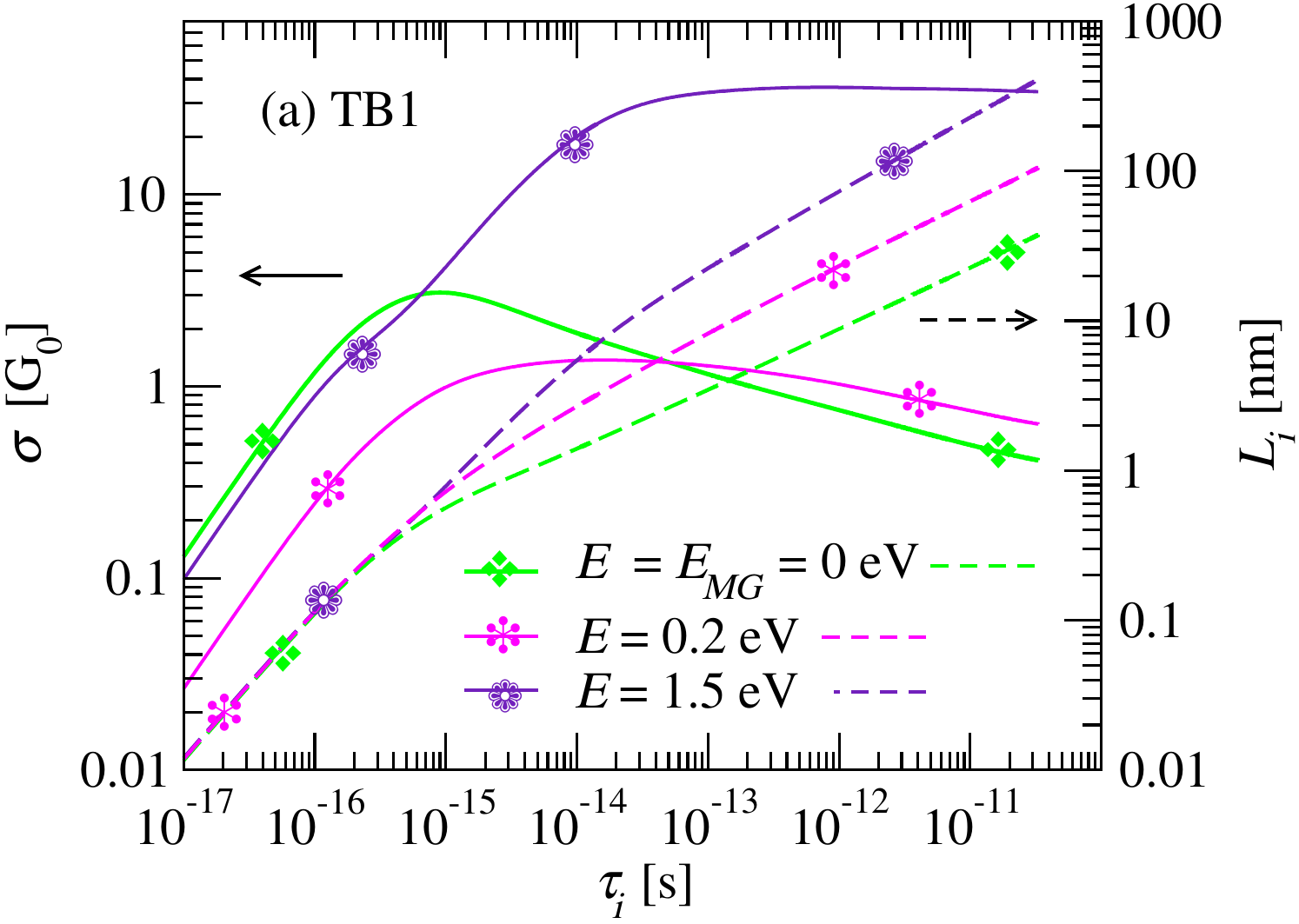}
}
\resizebox{0.43\textwidth}{!}{%
  \includegraphics{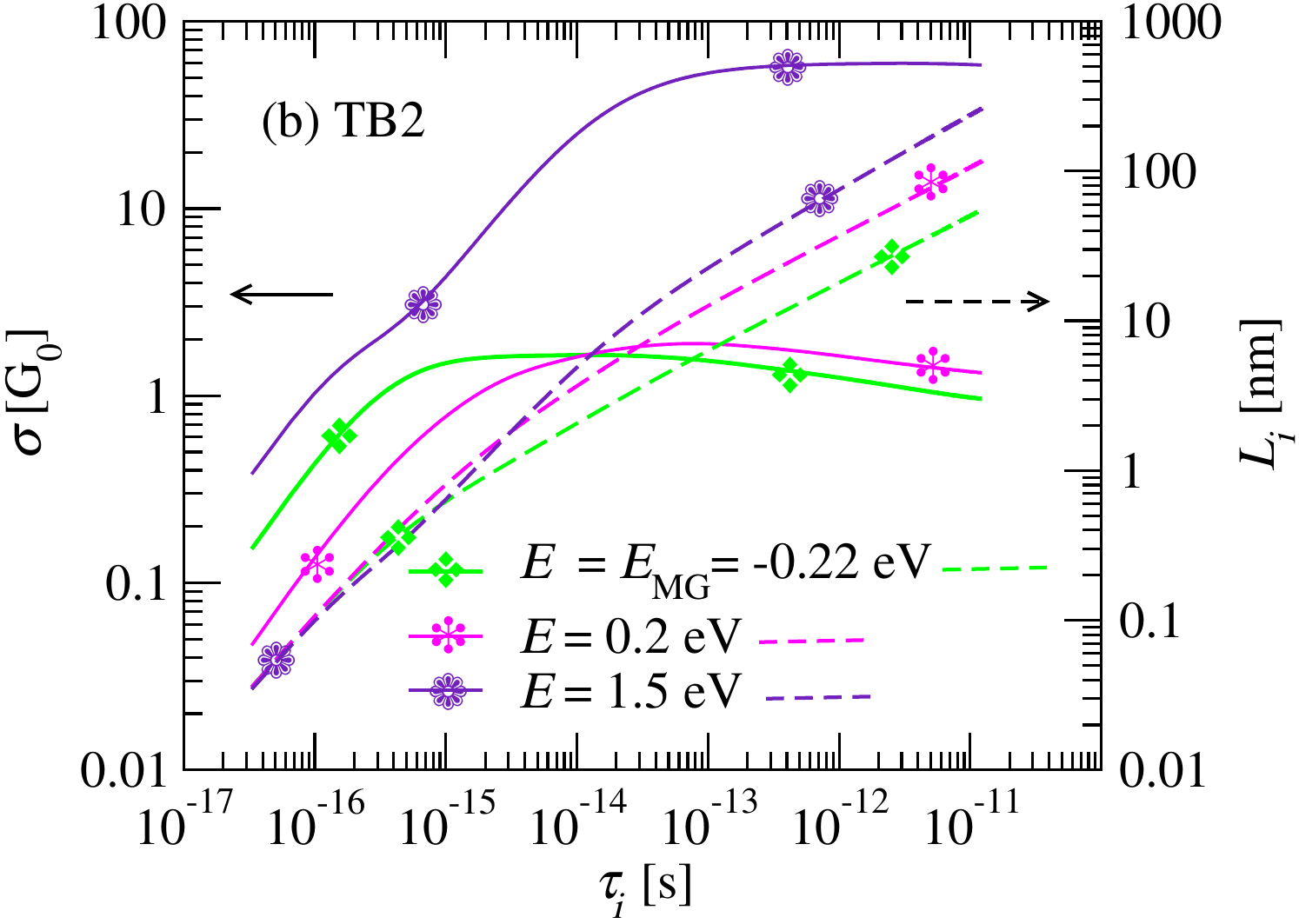}
}
\caption{\label{Fig3}
(color online)
Conductivity $\sigma$ (solid line) and inelastic scattering length $L_i$ (dashed line) in BLG versus inelastic scattering time $\tau_i$ for concentration $c = 1\%$ of resonant adsorbates (vacancies), and for 3 energy values: (a) TB1 (first neighbor hopping only), (b) TB2 (with hopping beyond first neighbors).
$G_0 = {2 e^2}/{h}$.
}
\end{center}
\end{figure}

We compute the distance $L_{i}$, the diffusivity $D$ and the conductivity $\sigma$ at all inelastic scattering times $\tau_i$ and all energies $E$ for model Hamiltonian that includes inelastic scatters distributed randomly in the super-cell.
At short times $\tau_i$ --$i.e.$ $\tau_i$ lower than elastic scattering time $\tau_{e}$-- the propagation is ballistic and the conductivity $\sigma$ increases when $\tau_{i}$ increases (Fig. \ref{Fig3}), 
\begin{equation}
L_i(E,\tau_i) \simeq V_0(E) \tau_i {\rm~ when~} \tau_i \ll \tau_e,
\end{equation}
where $V_0(E)$ is a velocity  at the energy $E$ and short time $t$.
In crystals,  $V_0 \ge V_B$ where $V_B$ is the Boltzmann velocity (intra-band velocity) \cite{Trambly14CRAS}.
In BLG and MLG, $V_0$ and $V_B$ have the same order of magnitude: 
$V_0({\rm BLG}) = V_0({\rm MLG}) = \sqrt{2}V_B({\rm MLG})$ \cite{Trambly16}.
According to the renormalization theory \cite{Lee} in 2D systems with static defects, diffusivity $D$ always goes to zero at very large $\tau_i$. 
At each energy, the microscopic diffusivity $D_M$ (microscopic conductivity $\sigma_M$) is defined as the maximum value of $D(\tau_i)$ ($\sigma(\tau_i)$).
We compute also the elastic mean-free path $L_{e}$ along the $x$-axis, from the relation \cite{Trambly13},
\begin{equation}
\label{le}
L_e(E) = \frac{1}{V_{0}(E)} Max_{\tau_i} \left\{ \frac{L_i^{2}(E,\tau_i)}{\tau_i} \right\} = \frac{2 D_M(E)}{V_{0}(E)}.
\label{Eq_Le}
\end{equation} 
$L_e$ is the average distance between two elastic scattering events. At each energy, the elastic scattering times $\tau_e$ is deduced from $L_e$ by $L_e(E) = V_0(E) \tau_e(E)$.

In our calculations $\tau_{i}$ and $L_i$  are considered as adjustable parameters.
Roughly speaking, when $L_i \ll L_e$ ($\tau_i \ll \tau_e$) the inelastic disorder dominates; it should correspond to very high temperatures. 
When $L_i \simeq L_e$ ($\tau_i \simeq \tau_e$), the conductivity is equal to microscopic conductivity, which should correspond to high temperature cases, typically room temperature. 
And when $L_i \gg L_e$ ($\tau_i \gg \tau_e$), quantum localization will dominates transport properties; this is the  low temperature limit. 
In the following we therefore discuss the two important cases: $L_i \simeq L_e$ ($\tau_i \simeq \tau_e$) and $L_i \ll L_e$ ($\tau_i \ll \tau_e$).

Figure \ref{Fig3} shows the variation of the conductivity $\sigma$ and the inelastic mean free path $L_i$ versus $\tau_{i}$ for energies corresponding to the three previous cases (see Sect. \ref{secDOS}).
The first case ($i$) ($E=1.5$\,eV in Fig. \ref{Fig3})
corresponding to a  Boltzmann transport: for large values of $\tau_{i}$, the conductivity $\sigma$ is almost constant as expected in a diffusive regime. This regime corresponds to energies for which the DOS is weakly affected by scatters.
The third case ($iii$) ($E=E_{\rm MG}$ in Fig. \ref{Fig3}) is determined by the transport of the midgap states which are localized states.  
The latter case ($ii$) is an intermediary regime between the previous two: 
for $\tau_{i}$ closed to the elastic scattering time $\tau_{e}$, there is a diffusive behavior where the $\sigma(\tau_i)$ reaches a maximum, $\sigma_M$; for larger values of $\tau_{i}$, $\tau_i \gg \tau_e$, $\sigma(\tau_i)$ decreases progressively as expected in localization regime due to Anderson localization in 2D \cite{Lee}.

In BLG another relevant time is the average traveling time $t_1$ between two interlayer hoppings of the charge carriers, which is associated to an
average traveling distance $l_1$ in a layer between two interlayer hoppings  \cite{McCann13,Snyman07}.
In perfect BLG typical values of $t_1$ and $l_1$ can be easily estimated : $t_1 = \hbar / \Gamma_1 \simeq 2 \times 10^{-15}$\,s, where $\Gamma_1 \simeq 0.4$eV is the 
interlayer hopping parameters of the Hamiltonian, and  
$l_1 \simeq V_m t_1 \simeq 2\, {\rm nm}$ 
{
\cite{Snyman07}
}
where $V_m$ is the velocity in MLG, 
$V_m \simeq  V_0 \simeq 10^6$\,ms$^{-1}$. 
When there is elastic disorder such that $ \tau_e < t_1 $ the value of $t_1$ can be modified. A simple argument may be given as follows: A Bloch state  of the  MLG is still coupled to Bloch states of the other layer by the same intensity, typically $\Gamma_1$, but these states are no longer eigenstates and have a typical lifetime $ \tau_e $. Because of that they have a spectral width $W\simeq  \hbar /\tau_e$. From the Fermi Golden rule the typical time needed to jump from one layer to the other will be such that  $\hbar/t'_1 \simeq \Gamma_1^2/W$. 
Therefore the new value of the interlayer hopping time $t'_1$ will be larger and will be such that  $t'_1\simeq t_1 (\hbar/\Gamma_1 \tau_e)$. Since the propagation is diffusive on the timescale $t>\tau_e$ with diffusion coefficient $D\simeq V_0^2 \tau_e$, the new length $l'_1$ in presence of defects is obtained from the relation, ${l'_1}^2/t_1' \simeq V_0^2 \tau_e$,
and thus $l_1' \simeq l_1 \simeq 2$\,nm depends weakly on disorder (i.e. $l_1'$ almost independent  on $\tau_e$). 

As shown in the following, electronic properties of disordered BLG depend on the values of the length $L_e$, $\xi$ and $l_1$ which are characteristic of the BLG and of the amount of elastic scatters:
\begin{itemize}
\item For low concentration of defects, $c < c_l = 1\%$-$2\%$, and for $E \ne E_{MG}$, we have, $\l_1 \le L_e \ll \xi$, and thus electronic  properties of BLG are influenced by  interlayer hoppings for every $L_i \ge L_e$ values. 
\item For larger concentration of defects, $c > c_l$, one obtains, $\L_e < l_1 \ll \xi$.  
When the effect of quantum interferences on conductivity is small, $i.e.$  when $L_e \simeq L_i < l_1$, BLG  behaves as two decoupled MLG. 
For $\L_e < l_1 \le L_i \ll \xi$, the coupling between the two planes can affect the propagation of charge carriers before inelastic scattering makes the propagation diffusive (i.e. on the length scale $L_i$). In this regime, quantum corrections to transport are not the same in the BLG and in two decoupled MLG. This influences strongly the localization regime as we discuss in Sect. \ref{secLocalization}. 
\end{itemize}


\begin{figure}
\begin{center}

\resizebox{0.45\textwidth}{!}{%
  \includegraphics{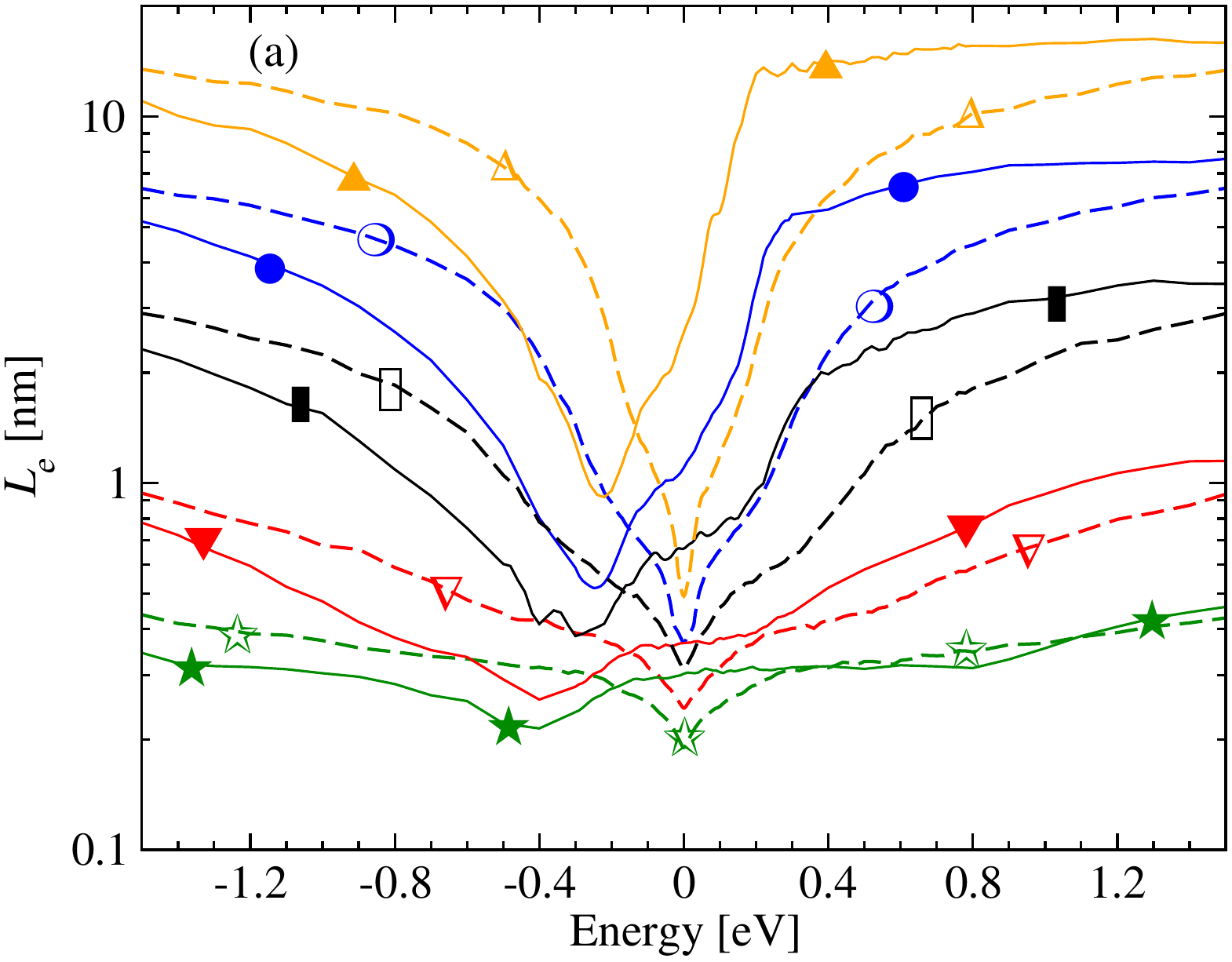}
}
\resizebox{0.45\textwidth}{!}{%
  \includegraphics{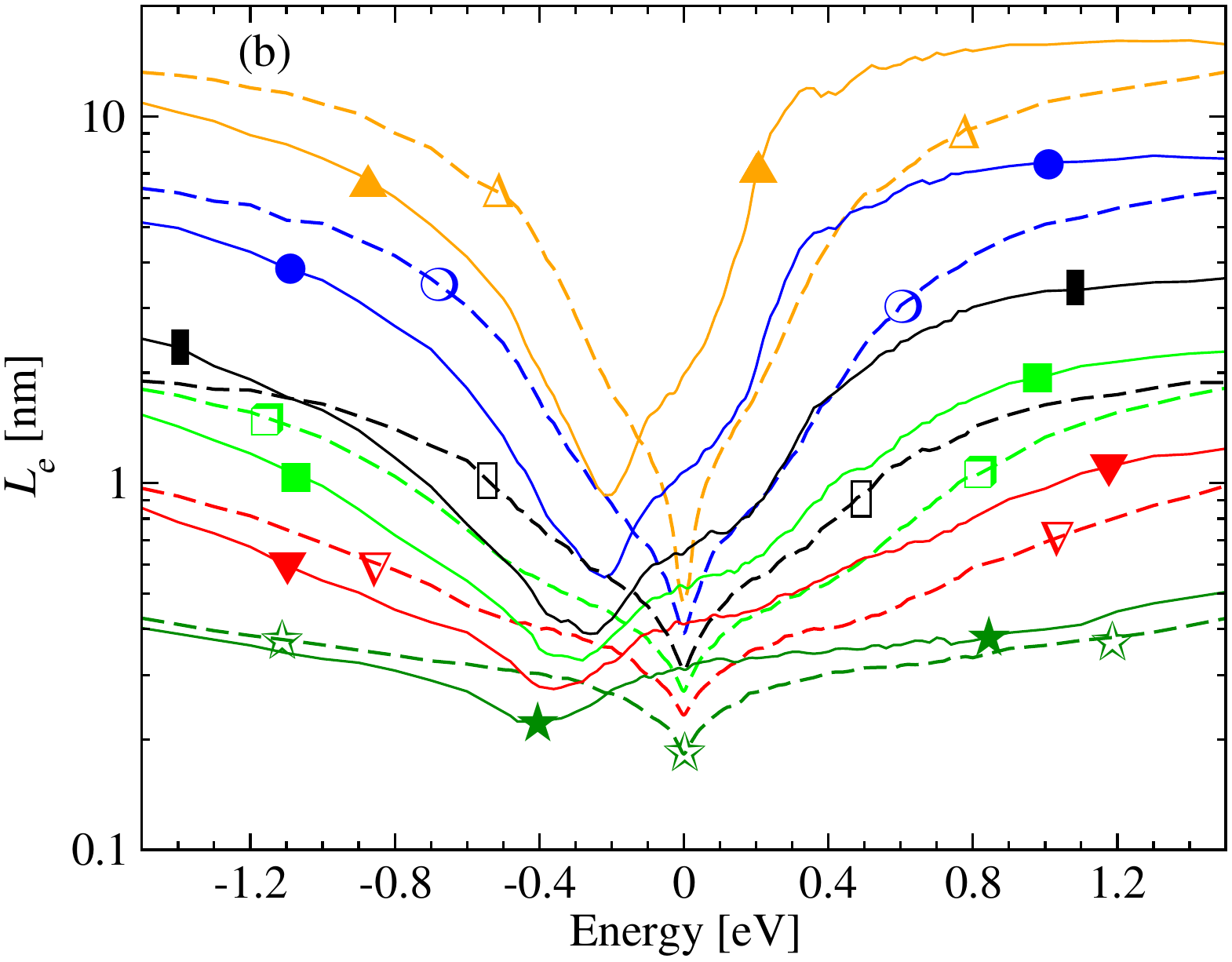}
}
\caption{\label{Fig_Le}
(color online)
Elastic mean free path $L_e(E)$ versus energy $E$ (a) for MLG, (b) for BLG: 
(dashed line) TB1 (first neighbor hopping only), (solid line) TB2 (with hopping beyond first neighbor). 
Concentration of vacancies: triangle up $c=0.5\%$, circle $c=1\%$, rectangle $c=2\%$, square $c=3\%$, triangle down $c=5\%$, star $c=10\%$. }
\end{center}
\end{figure}

\subsection{Elastic mean-free path}
\label{secLe}

The elastic mean-free path $L_e$ (Eq. (\ref{Eq_Le})) along the $x$-axis 
as a function of the $E$ is shown figure \ref{Fig_Le} for different values of vacancy concentrations $c$ in MLG and BLG with both TB models.
It depends on the energy even in the intermediate regime and it takes a finite and non-zero value for $E=E_{D}$ but stays comparable to the distance $d$ between adsorbates (vacancies) defined by, $d \simeq {1}/{\sqrt{n_a}}$, 
where $n_a$ is the adsorbates density.
Numerical results (Fig. \ref{Fig_Le}) show that $L_e$ values in BLG and MLG 
{
are close to each other.
}
Moreover,
$L_e(E) < l_1 \simeq 2 $\,nm ($i.e.$ $\tau_e < t_1$)
for $c > c_l \simeq 1\%$--$2\%$; 
whereas  for smaller $c$, $L_e(E) \ge l_1$ ($i.e.$ $\tau_e \ge t_1$)
for $c < c_l$.

\subsection{Microscopic conductivity}
\label{secSigmaM}

\begin{figure}
\begin{center}
\resizebox{0.45\textwidth}{!}{%
\includegraphics{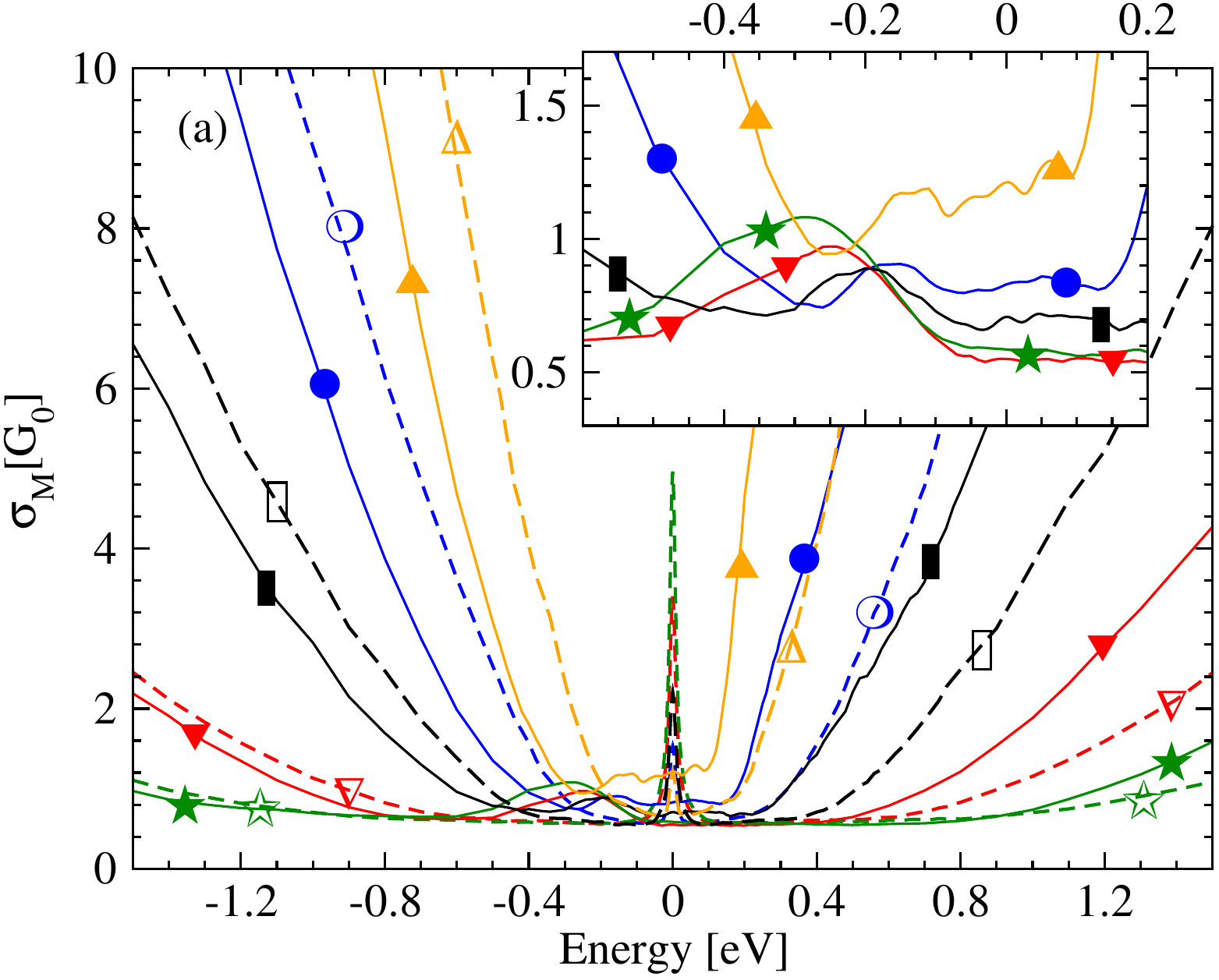} 
}

\resizebox{0.45\textwidth}{!}{%
\includegraphics{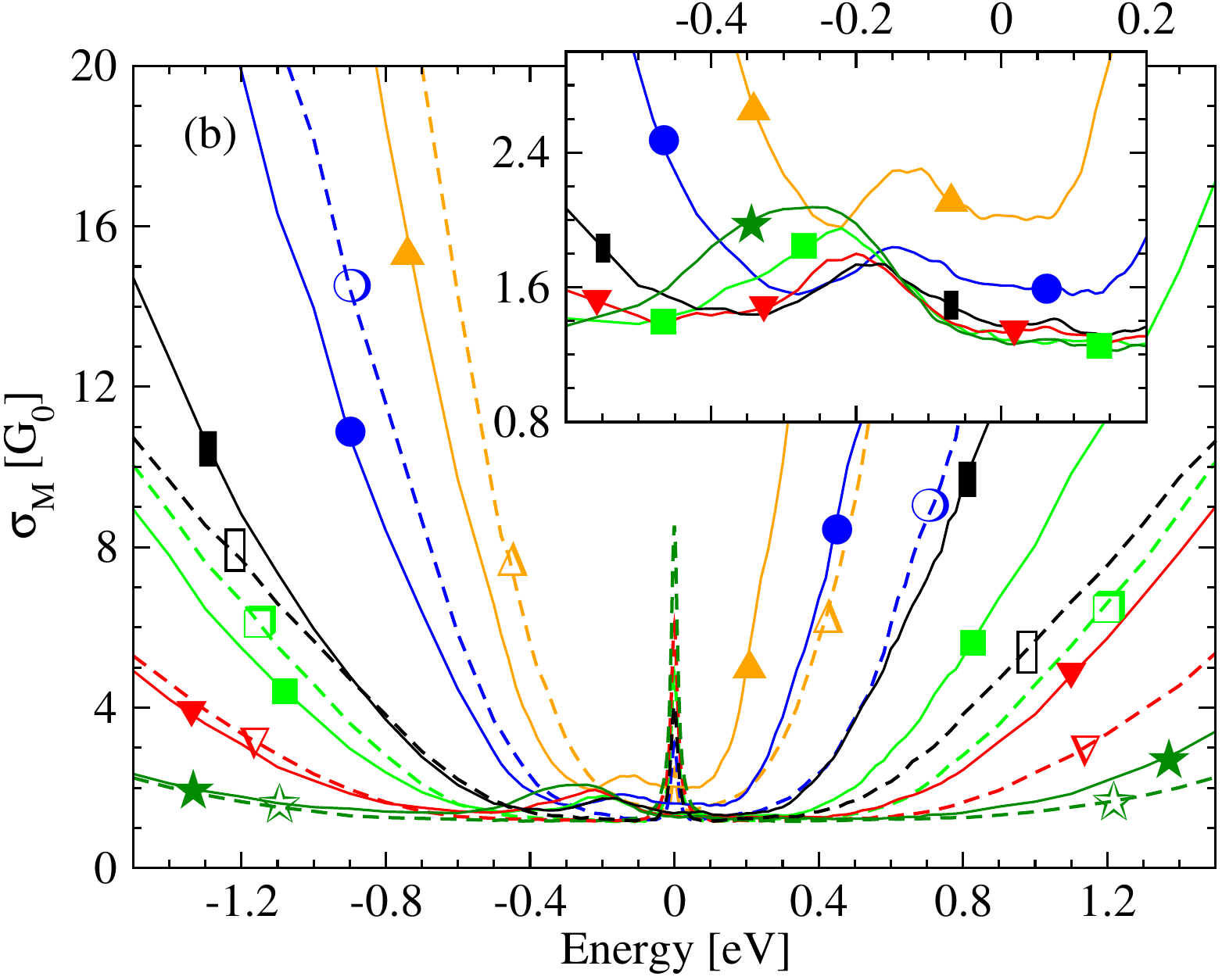}
}
\caption{\label{Fig_sigmaM}
(color online)
Microscopic conductivity $\sigma_M$ versus energy $E$ for (a) MLG and (b) BLG:  
(dashed line) TB1 (first neighbor hopping only), (solid line) TB2 (with hopping beyond first neighbor).
Concentration of vacancies: triangle up $c=0.5\%$, circle $c=1\%$, rectangle $c=2\%$, square $c=3\%$, triangle down $c=5\%$, star $c=10\%$.
{
Insert shows enlarged curves around $E_D$ for TB2.
}$G_0 = {2 e^2}/{h}$.}
\end{center}
\end{figure}

\begin{figure} 
\begin{center}
\resizebox{0.43\textwidth}{!}{%
\includegraphics{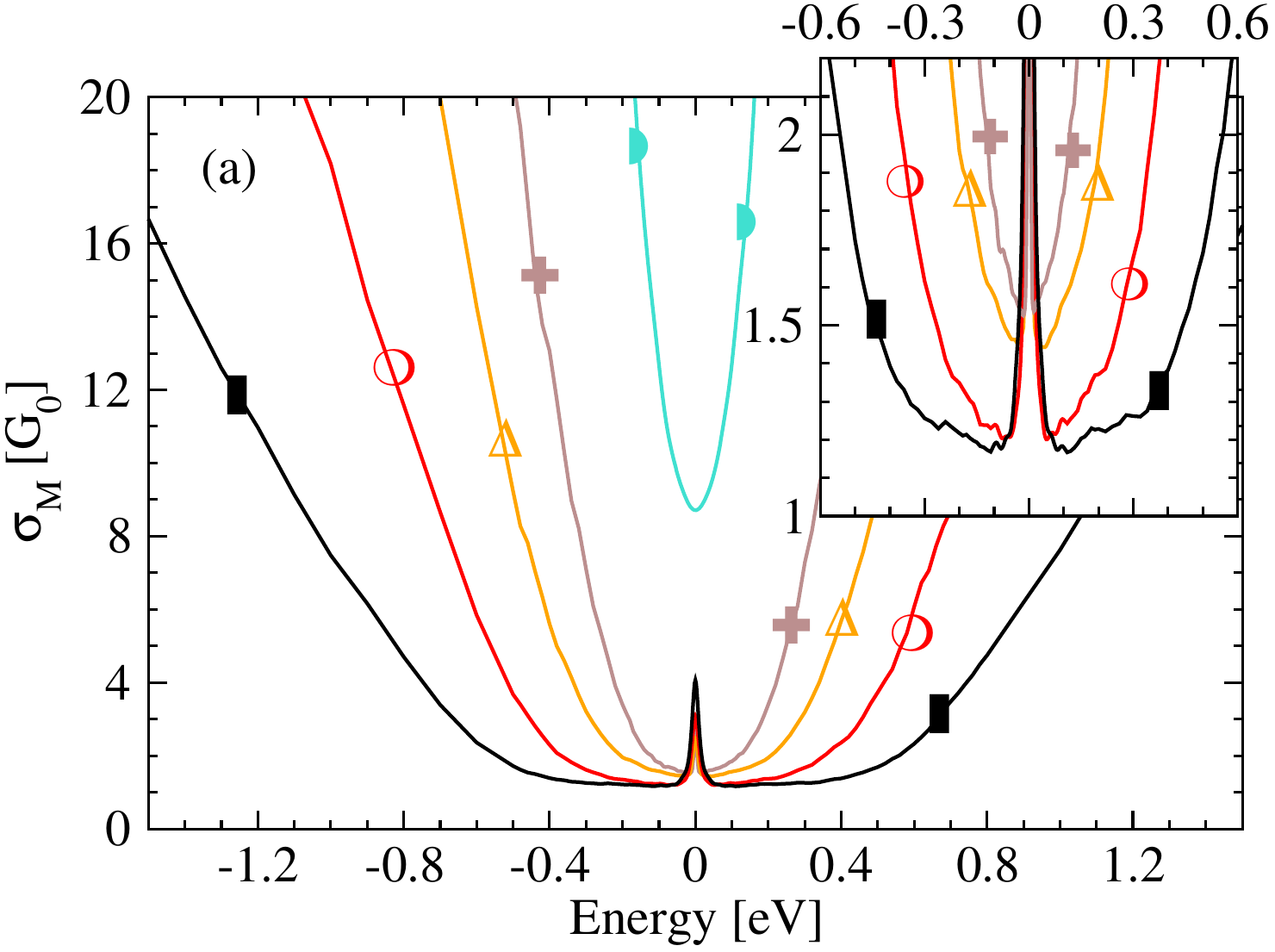}
}

\resizebox{0.43\textwidth}{!}{%
\includegraphics{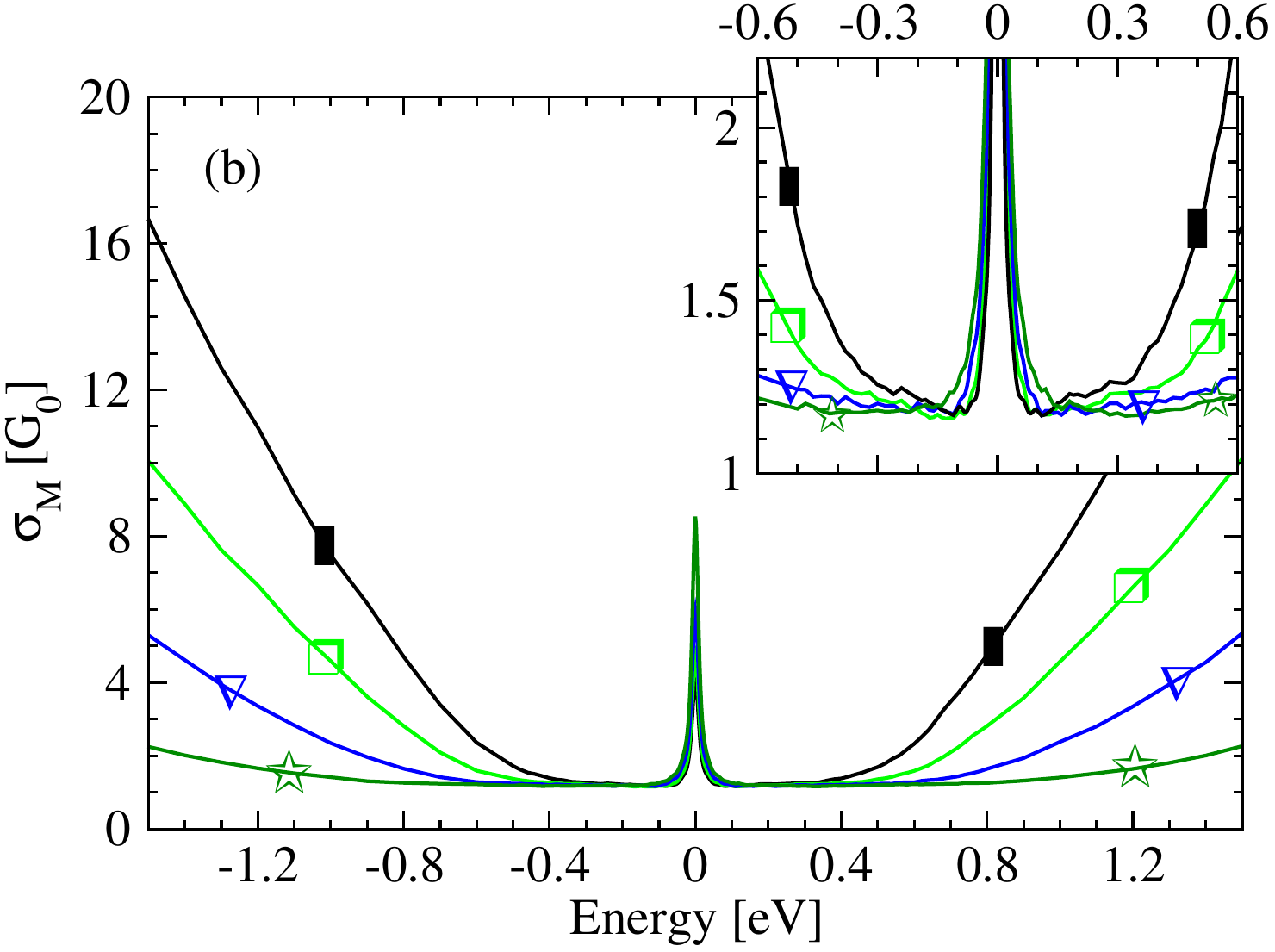}
}

\caption{\label{Fig_sigmaM2} 
(color online)
Microscopic conductivity $\sigma_M$ in BLG 
for TB1 (first neighbor hopping only): 
{
(a) $c<c_l$ and $\sigma_{M}$ with a constant minimum value (“plateau”) which decreases as $c$ increases, (b) $c>c_l$ and $\sigma_{M}$ reaches a minimum value independent of $c$ value (see text).
Concentration of vacancies: semi circle $c=0.05\%$, cross $c=0.25\%$,  triangle up $c=0.5\%$, circle $c=1\%$, rectangle $c=2\%$, square $c=3\%$, triangle down $c=5\%$, star $c=10\%$.
Insert shows enlarged curves around $E_D=0$.
}$G_0 = {2 e^2}/{h}$.
}
\end{center}
\end{figure}

As shown Fig. \ref{Fig3}, the static scattering events perturb strongly the wave packet propagation and a maximum value of the conductivity $\sigma(\tau_{i})$, called microscopic conductivity $\sigma_{M}$, $\sigma_{M}(E) = e^{2}n(E)D_M(E)$, is reached. 
$\sigma_{M}$ calculated from both TB models are shown in Figs. \ref{Fig_sigmaM} and \ref{Fig_sigmaM2} for different  concentrations $c$ of vacancies  in MLG and BLG.
According to the renormalization theory \cite{Lee}, this value is obtained when the inelastic mean free path $L_{i}$ and the elastic mean free path $L_{e}$ are comparable, $L_e \simeq L_i$, which corresponds to $\tau_i \simeq \tau_e$. 
As $L_i$ and $\tau_i$ decrease when the temperature $T$ increases, the microscopic conductivity is a good estimation of the high temperature conductivity (or room temperature conductivity).  

For energies corresponding the to Boltzmann regime, $i.e.$ regime ($i$) described in Sec. \ref{secDOS}, $\sigma_{M}\simeq \sigma_{B}$, where $\sigma_{B}$ is calculated with the Bloch-Boltzmann approach \cite{Castro09_RevModPhys,McCann13}. In this regime the conductivity decreases with the concentration of defects. 

In the intermediate energy values regime ($ii$), the semi-classical approach fails and the behavior depends on $c$. 
From 
{
Fig. \ref{Fig_sigmaM2}(a),
}
for small $c$ values, typically 
$c< c_l \simeq 1\%$--$2\%$ ($i.e.$ $L_i \simeq L_e \ge l_1$), 
$\sigma_{M}$ seems to reach a constant minimum value (``plateau''), but this minimum  $\sigma_{M}$ value decreases as $c$  increases. 
{
This concentration dependence is specific to BLG and is not observed in MLG.  
}
For larger $c$ 
{
(Fig. \ref{Fig_sigmaM2}(b)), 
}
$c>c_l$ ($i.e.$ $L_i \simeq L_e < l_1$),
$\sigma_{M}$ reaches a minimum value independent on $c$ value: 
$\sigma_{M} \simeq 1.2 G_0$ where $G_0 = {2 e^2}/{h}$. 
This values for BLG is \cite{Yuan10,Gonzalez10} two times the universal value of the conductivity, $\sim 4{e^{2}}/{(\pi h)}$,
expected 
{
in presence of resonant scatters in MLG \cite{Yuan10b,Lherbier12,Roche13,Trambly11,Trambly13,Zhao15}.
}

Results with TB2 model (including hopping beyond nearest neighbors) 
{
(fig. \ref{Fig_sigmaM})
}
show that a plateau of the microscopic  conductivity near the Dirac energy exists 
{
in MLG and BLG,
}
but is not symmetric due to the symmetry breaking electron-hole. 
Nevertheless in this case, 
{
for $c>c_l$, 
}
$\sigma_{M}$ values are still close to the universal conductivity plateau;
{
whereas for $c<c_l$, the minimum value of $\sigma_{M}$ decreases as $c$  increases like with TB1 model.
}

For energies in the midgap states ($iii$) with TB1 model (first neighbor hopping only),
an anomalous behavior of the conductivity is obtained and there is a peak of $\sigma_M$ at $E=E_{\rm MG}=0$. 
With TB2 (including hopping beyond nearest neighbors), this anomalous behavior is still slightly present at 
$E \simeq E_{\rm  MG}$, but the change in the conductivity is rather small. 
Thus, as in monolayer graphene \cite{Trambly11}, conduction by the midgap states is very specific to TB1 model. 

\subsection{Quantum localization regime}
\label{secLocalization}

\begin{figure} 
\resizebox{0.5\textwidth}{!}{%
\includegraphics{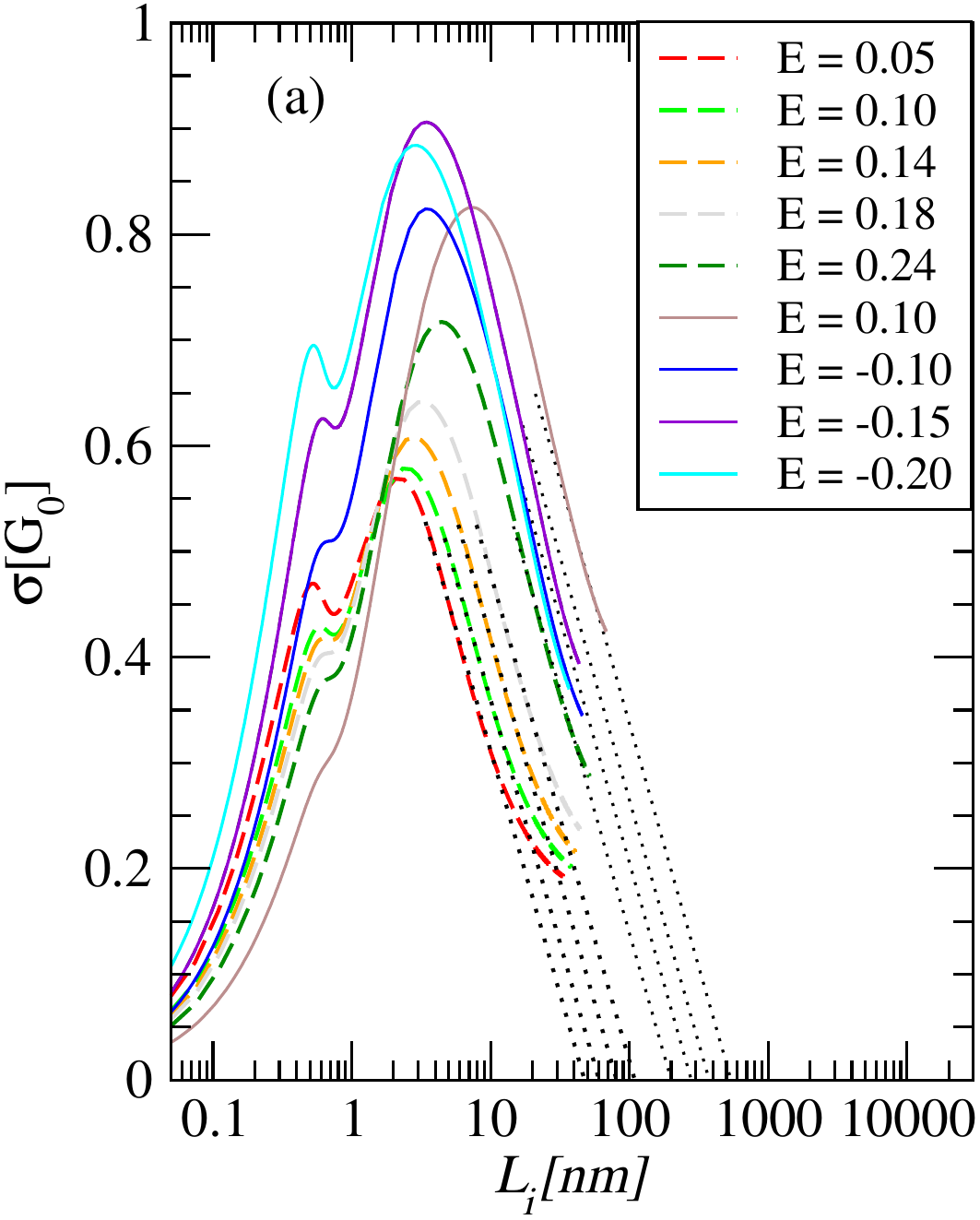}
\includegraphics{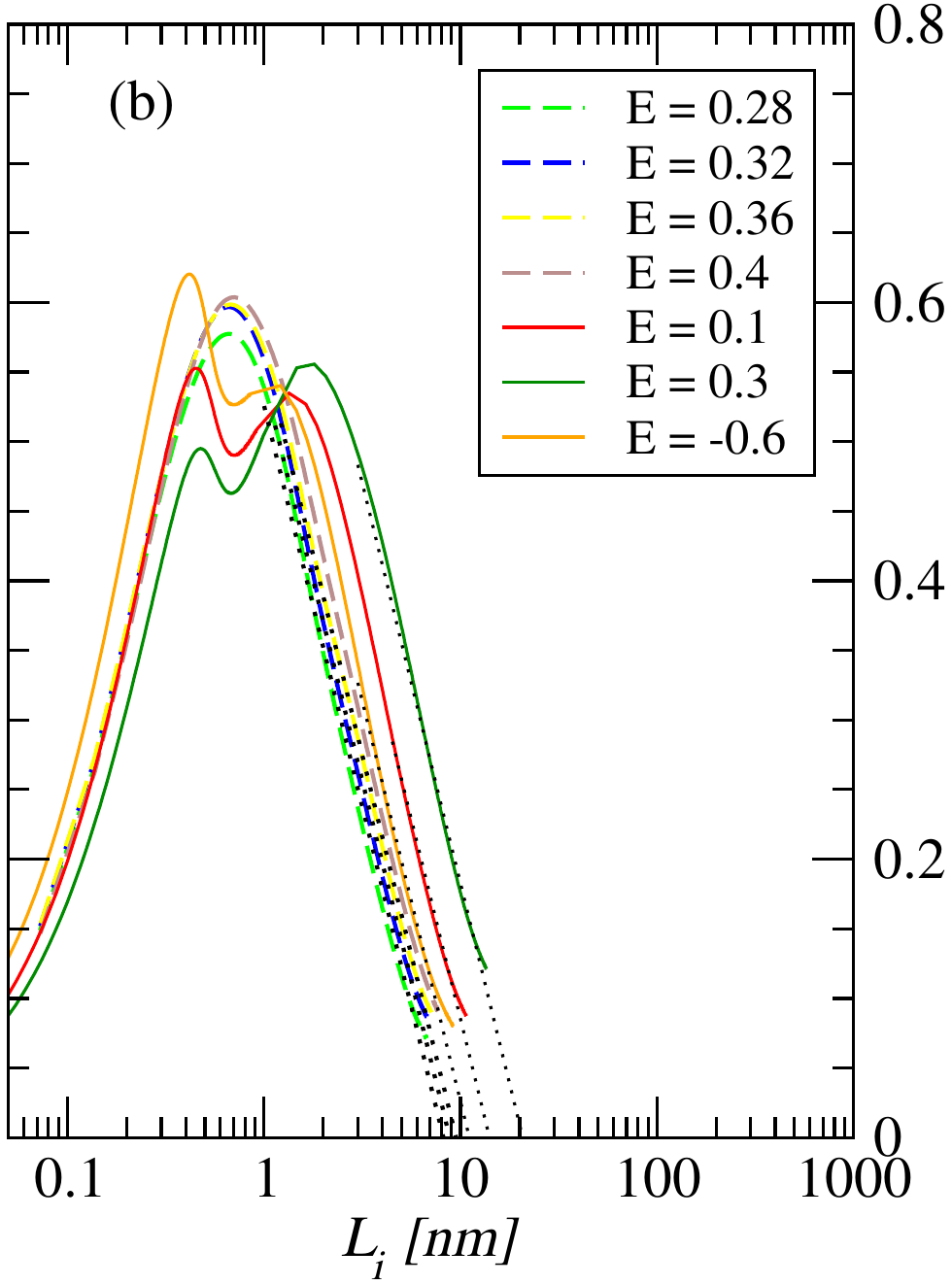}
}
\resizebox{0.5\textwidth}{!}{
\includegraphics{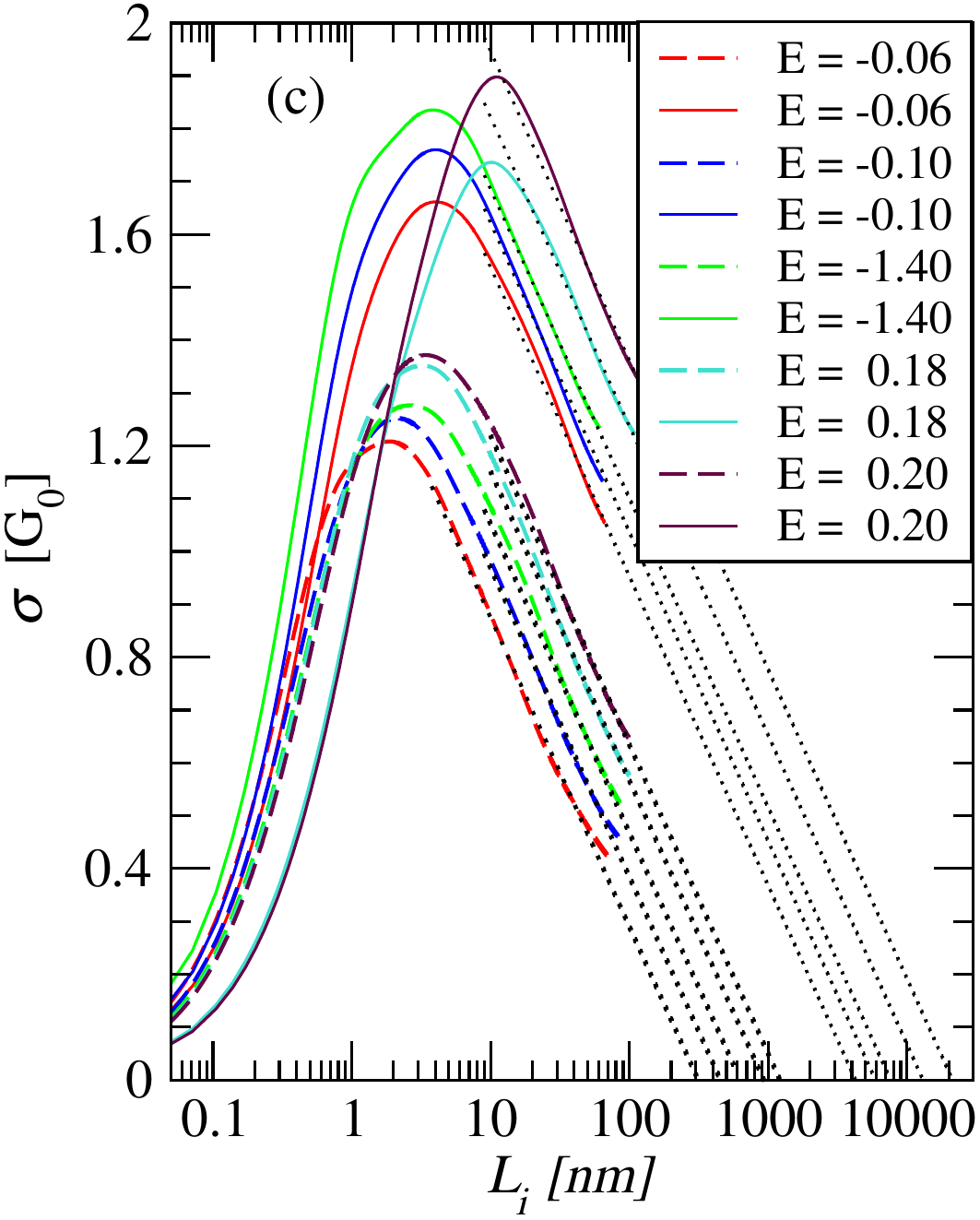}
\includegraphics{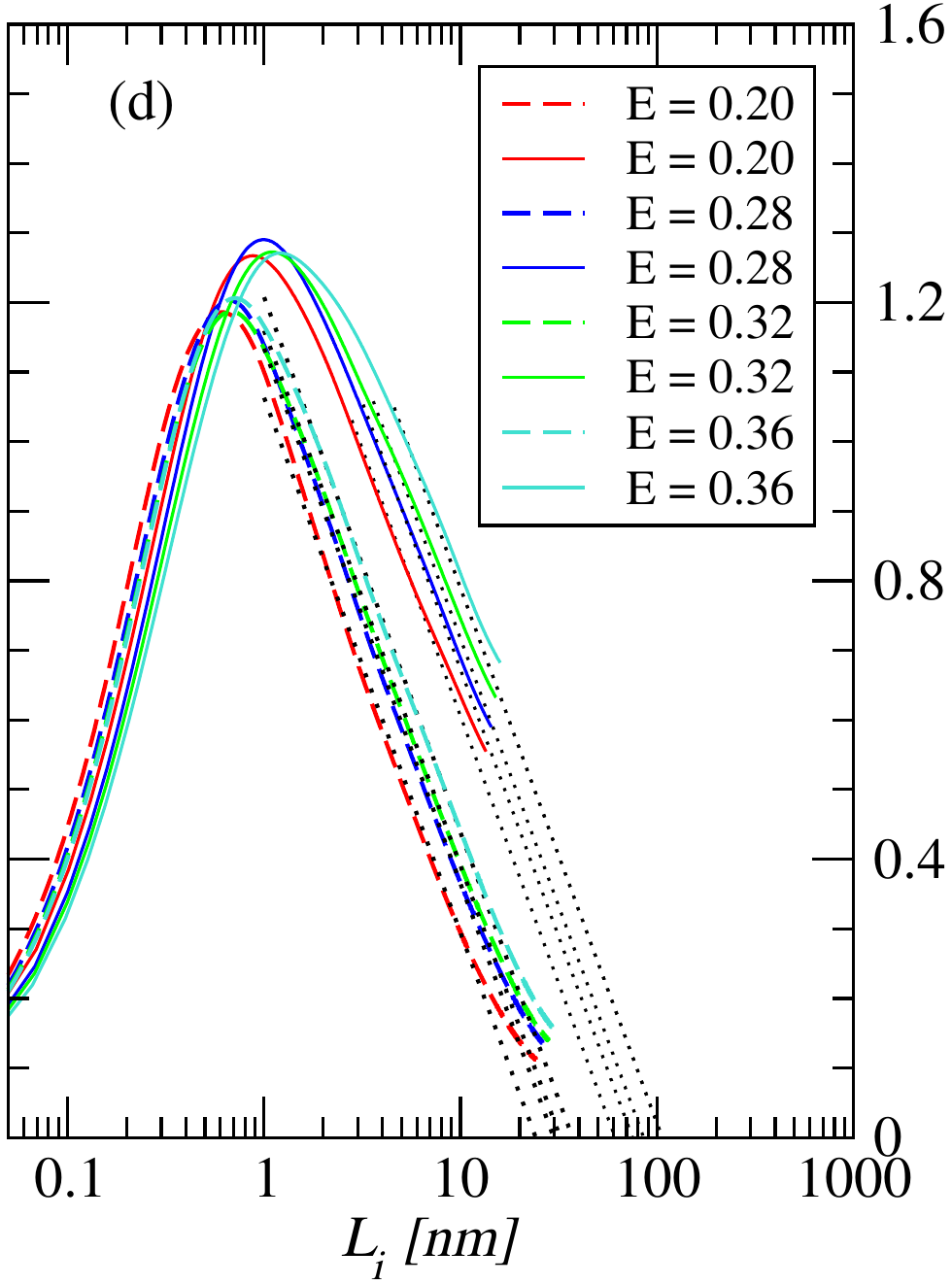}
}
\vspace{.2cm}
\caption{\label{Fig7}
(color online)
Conductivity $\sigma(L_i)$ versus the inelastic scattering length $L_{i}$ in (a) (b) MLG and (c) (d) BLG, for concentration (a) (c) 1\% and  (b) (d) 5\%, at different energies $E$ [eV] in the plateau of $\sigma_M(E)$: (dashed line) TB1 (first neighbor hopping only),  (solid line) TB2 (with hopping beyond first neighbor).  
$G_0 = {2 e^2}/{h}$.
}

\end{figure}

\begin{figure} 
\begin{center}
\resizebox{0.36\textwidth}{!}{%
\includegraphics{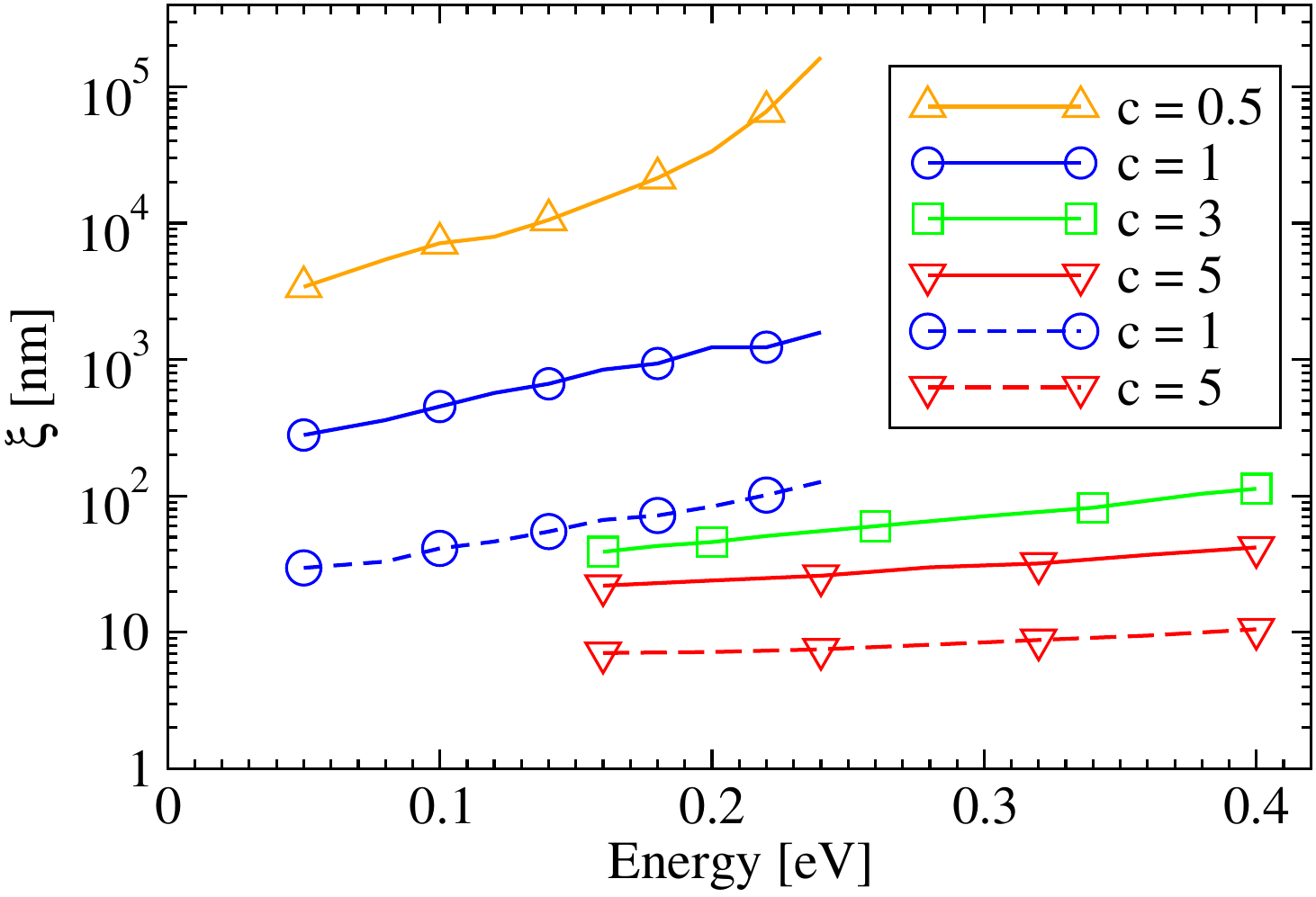}
}
\caption{
(color online)
\label{Fig8} 
Localization length versus energy for different concentrations $c$ [\%] of vacancies: (dashed line) MLG, (solid line) BLG.
This results are computed with TB1 model (first neighbor hopping only).
}
\end{center}
\end{figure}

In this section we consider the case of large inelastic mean-free path, 
$L_i \gg L_e$ ($i.e.$ $\tau_i \gg \tau_e$), 
which should corresponds to the  low temperature limit.
The conductivity in the intermediate energy case ($ii$) (see Sect. \ref{secDOS}) presents localization effects that  are a consequence of quantum interferences. 
In that case, for all vacancy concentration values $c$, $L_i \gg l_1$ ($i.e.$ $\tau_i \gg t_1$). 
Therefore BLG should have different localization behavior than MLG.
As shown in Fig. \ref{Fig7}, $\sigma(L_i)$ follows the linear variation with the logarithm of the inelastic mean free path $L_i$, 
like in the case of monolayer graphene \cite{Trambly11,Trambly13},
\begin{equation}
\label{localisation}
\sigma(E,L_i) = \sigma_0(E) - \alpha G_0 \log \left(\frac{L_i}{L_e(E)} \right),
\label{equationSig_FLi}
\end{equation}
where $G_0 = {2 e^2}/{h}$, $L_e(E)$ is defined by Eq. (\ref{Eq_Le}), and $\sigma_0$ values are in the range of $\sigma_M$ values. 
For low concentration of defects $c=0.5\%$ and $1\%$ one can estimate $\alpha \simeq 0.26$, and for larger $c$, $c = 3\%$ and $5\%$,
$\alpha \simeq 0.32$.
These values are close to the result found in monolayer graphene with same computational method \cite{Trambly13} and close to the prediction of the perturbation theory of 2D Anderson localization for which $\alpha = 1/\pi$ \cite{Lee}. 
As in the monolayer case, this linear variation of $\sigma$ with $\log L_i$ is found for both models, with nearest neighbor hopping only and with hopping beyond first nearest neighbors.
 
We finally define the localization $\xi$ length from the expression (\ref{equationSig_FLi}) by extrapolation of $\sigma(L_i)$ curves  (Fig. \ref{Fig7}) when : $\sigma(L_i = \xi)=0$, and then 
\begin{equation}
\xi(E) = L_e(E) \exp \left( \frac{\sigma_0(E)}{\alpha G_0} \right).
\end{equation}
The $\xi$ values for energies in the plateau of $\sigma_M$ ($i.e.$ case (ii) described in Sect. \ref{secDOS})  are shown figure \ref{Fig8}. 
For large concentrations of defects ($c>c_l$), $\xi$ in MLG and BLG is almost independent on the energy. 
Moreover $\xi$(BLG) is always larger than $\xi$(MLG). 
{
This difference results from the fact that $\sigma_{0, BLG} \simeq 2 \sigma_{0, MLG}$ (Sect. \ref{secSigmaM}) and then, for $c>c_l$,
\begin{equation}
\left ( \frac{\xi(E)}{L_e(E)} \right )_{BLG} \simeq \left ( \frac{\xi(E)}{L_e(E)} \right )_{MLG}^2 ,
\end{equation}
where $L_e$ are similar in MLG and BLG (Sec. \ref{secLe}). 
It is thus a multilayer effect on quantum interferences that modifies the 2D behavior with respect to MLG cases.  
For low defects concentration ($c<c_l$), interlayer coupling modifies also quantum interferences.
}
Therefore,  for every resonant scatterer concentration,  quantum interference corrections to the conductivity in BLG and in MLG  are not similar.

\section{Conclusion}
To conclude we have studied numerically the quantum diffusion of charge carriers in monolayer graphene and Bernal bilayer graphene in the presence of local defects. These defects are simulated by simple vacancies  randomly distributed in the structure. Among the fundamental length scales in the MLG and BLG  there are the elastic mean free path  $L_{e}$, the localization length $\xi$ and the  inelastic mean-free path $L_{i}$ which in real systems depends on temperature or on magnetic field. For the bilayer, there is an additional length scale which is the  typical distance $l_1\simeq 2$\,nm over which an electron travels in a plane before hopping to the other plane. We have compared the Bilayer and monolayer transport properties for identical concentrations of vacancies. We show that these properties can be either similar or different depending on the  comparison between $l_1$ and the other three length scales $L_{e}$, $L_{i}$, $\xi$. This relation explains essentially the numerical results detailed in this paper.

Our results show that for strong concentration of defects, $c > c_l \simeq 1\%$--$2\%$, the bilayer graphene could be equivalent to two independent disordered monolayers of graphene, because the elastic mean free path $L_e$ is smaller than the average distance $l_1$. 
Therefore for $c > c_l$, the universal aspects of the conductivity  are present in bilayer, as in monolayer graphene, with (TB1) or without (TB2) the hopping beyond nearest neighbors.
 
In the high temperature limit, $i.e$ when inelastic scattering length $L_i$ is small, $L_i \simeq L_e$, the conductivity in bilayer  
{
is almost equal to two times the universal minimum plateau
of microscopic conductivity in monolayer graphene 
}
(except for the Dirac energy with TB1 model that takes only into account nearest neighbor hopping). 
{
For smaller $c$, $c < c_l$, the BLG should be considered like a usual metal:  with static defects the minimum of microscopic conductivity of BLG increases when the $c$ values decreases.
}
For the parameters studied here, the localization length $\xi$ is larger than the traveling distance $l_1$ between two interlayer hopping; and therefore the BLG and MLG have different localization lengths at the same concentration, even if they have similar elastic-mean-free paths. The localization length is the largest in the BLG. In the limit, $i.e$ $L_i \gg L_e$, (which is relevant at low temperature) and for all $c$ values,
the conductivity follows a linear variation with the logarithm of $L_i$ in MLG and BLG and for both TB1 models (nearest neighbor hopping only) and TB2 (with hopping beyond nearest neighbors), excepted for energy in the midgap states for TB1. This is in good agreement with two-dimensional Anderson localization and consistent with the expected universal behavior of conductivity of a two dimensional disordered system \cite{Lee}.

\section*{Acknowledgment}
The authors wish to thank C. Berger, W. A. de Heer, L. Magaud, P. Mallet and J.-Y. Veuillen
for fruitful discussions.
The numerical calculations have been performed 
at  Institut N\'eel, Grenoble,
and at the Centre de Calculs (CDC),
Universit\'e de Cergy-Pontoise.
This work was supported by the Tunisian French Cooperation Project (Grant No. CMCU 15G1306)
and the project ANR-15-CE24-0017.

{
\section*{Author contribution statement}
Numerical calculations have been performed by A. Missaoui and G. Trambly de Laissardi\`ere. 
All authors took part in drafting of the manuscript as well as in analysis and interpretation of the
results.
}


%


%
%

\end{document}